\documentclass[
  ncsstyle   = false,
  removetemp = true
]{ncs-ieee-conference}

\IEEEoverridecommandlockouts                              

\AtBeginBibliography{\footnotesize}

\AtEveryBibitem{%
  \clearfield{note}%
  \clearfield{url}%
  \clearfield{urldate}%
  \clearfield{urlyear}%
}

\begin{document}

\title{\LARGE \bf
  \ncsblue{
    Attainable Force Approximation and Full-Pose Tracking Control of an Over-Actuated Thrust-Vectoring Modular Team UAV}%
}

\author{Yen-Cheng Chu, Kai-Cheng Fang, and Feng-Li Lian
\thanks{The authors are with National Taiwan University, Taipei, Taiwan. (email: {\tt\small fengli@ntu.edu.tw}). %
This research is supported by the Ministry of Science and Technology, Taiwan, under Grants: MOST 110-2221-E-002-168 and 111-2221-E-002-191.}%
}

\maketitle
\thispagestyle{empty}
\pagestyle{empty}

\begin{abstract}
  Traditional vertical take-off and landing (VTOL) aircraft can not achieve optimal efficiency for various payload weights and has limited mobility due to its under-actuation.
  With the thrust-vectoring mechanism, the proposed modular team UAV is fully actuated at certain attitudes.
  However, the attainable force space (AFS) differs according to the team configuration, which makes the controller design difficult.
  We propose an approximation to the AFS and a full-pose tracking controller with an attitude planner and a force projection, which guarantees the control force is feasible.
  %
  %
  %
  The proposed approach can be applied to UAVs having multiple thrust-vectoring effectors with homogeneous agents.
  The simulation and experiment demonstrate a tilting motion during hovering for a 4-agent team.
\end{abstract}


\newcommand{\BiRAi}{\,^{\mathcal{B}_i}{\mathbf{R}}_{\mathcal{A}_i}}
\newcommand{\BidRAi}{\,^{\mathcal{B}_i}\dot{\mathbf{R}}_{\mathcal{A}_i}}
\newcommand{\BRA}{\,^\mathcal{B}\mathbf{R}_{\mathcal{A}}} %
\newcommand{\IRB}{\,^{\mathcal{I}}{\mathbf{R}}_{\mathcal{B}}}
\newcommand{\FVBi}[4]{\,^{\mathcal{#1}}{\mathbf{#2}}_{\mathcal{#3}_{#4}}}
\newcommand{\dFVBi}[4]{\,^{\mathcal{#1}}\dot{{\mathbf{#2}}}_{\mathcal{#3}_{#4}}}
\newcommand{\FVB}[3]{\FVBi{#1}{#2}{#3}{}}
\newcommand{\dFVB}[3]{\dFVBi{#1}{#2}{#3}{}}
\newcommand{\BRP}[1]{\,^\mathcal{B}\mathbf{R}_{\mathcal{P}_{#1}}} %

\newcommand{\FVi}[3]{\,^{\mathcal{#1}}{\mathbf{#2}}_{#3}}
\newcommand{\dFVi}[3]{\,^{\mathcal{#1}}\dot{{\mathbf{#2}}}_{#3}}

\section{Introduction}

Using UAVs for transportation or delivery has become a popular topic, especially for vertical take-off and landing (VTOL) UAVs. 
VTOL has numerous advantages such as long-range, take-off and landing without a runway, and the capability to change flight modes depending on task requirements. 
In a typical transportation scenario, payloads usually vary in size and weight. However, an aircraft is limited by the maximum take-off weight or becomes a trade-off between flight range and payload.
%
Using multiple UAVs to enhance insufficient mobility or capacity is a highly discussed method in research. 


The benefits of modularity do not grow with scalability due to the saturation of individual agents and the fast-growing team inertia \cite{gabrichModQuadDoFNovelYaw2020}, 
which can be solved using tilted rotors \cite{xuHModQuadModularMultiRotors2021}. 
However, it has unavoidable power efficiency drops due to the cancellation of internal forces.
Although work like \cite{ryllModelingControlFASTHex2016} can alter the tilting angle to adjust the degree of power loss, however, it is not designed to respond in real-time.
Thrust-vectoring vehicles using rotatable thrusters \cite{orrHighEfficiencyThrustVector2014, kamelVoliroOmniorientationalHexacopter2018} and control the deflection angles in the loop.
Passively-actuated systems use smaller UAVs as passive thrusters to perform dexterous tasks.
Multiple quadrotors are connected using spherical joints, forming unconstrained \cite{suDownwashawareControlAllocation2022a} or bundled cone constrained \cite{nguyenNovelRoboticPlatform2018} pointing motion on agents.
The full-pose tracking problem for laterally-bounded force systems is discussed in \cite{franchiFullPoseTrackingControl2018}, 
but agents are usually bounded by maximum propeller speeds, forming a cone-shaped attainable force space (AFS) \cite{suNullspaceBasedControlAllocation2021, nguyenNovelRoboticPlatform2018}.
%
\cite{invernizziTrajectoryTrackingControl2018} describes the attainable force space of a UAV as a magnitude-bounded spherical sector, and a temporary attitude target is designed accordingly to relax an infeasible reference.
Furthermore, the nested saturations-based nonlinear feedback \cite{naldiRobustGlobalTrajectory2017} is integrated \cite{invernizziDynamicAttitudePlanning2020}, and the solutions are proven to be almost globally asymptotic stable.
Methods in \cite{franchiFullPoseTrackingControl2018, nguyenNovelRoboticPlatform2018, invernizziDynamicAttitudePlanning2020} achieve full-pose tracking controls by designing suitable attitude planners based on the system's attainable force space, however, they can not cover all the attainable space in a system having elliptic cone constraints.
The contributions are summarized as follows:
\begin{itemize}
    \item We propose a novel design of a reconfigurable modular team UAV that solves the actuation problem of aerial team systems and provides high mobility and reliability simultaneously.
    \item By introducing an online attainable force space approximation, the attitude planner is capable of generating feasible attitude targets for the team UAV, such that it can be applied to various team configurations without the need to redesign controller parameters.
\end{itemize}




\section{Module Design and The Team System}
\begin{figure}[t]
    \centering
    \includegraphics[width=1.0\linewidth, trim={{1.0cm} {0cm} {1.5cm} {0cm}}, clip]{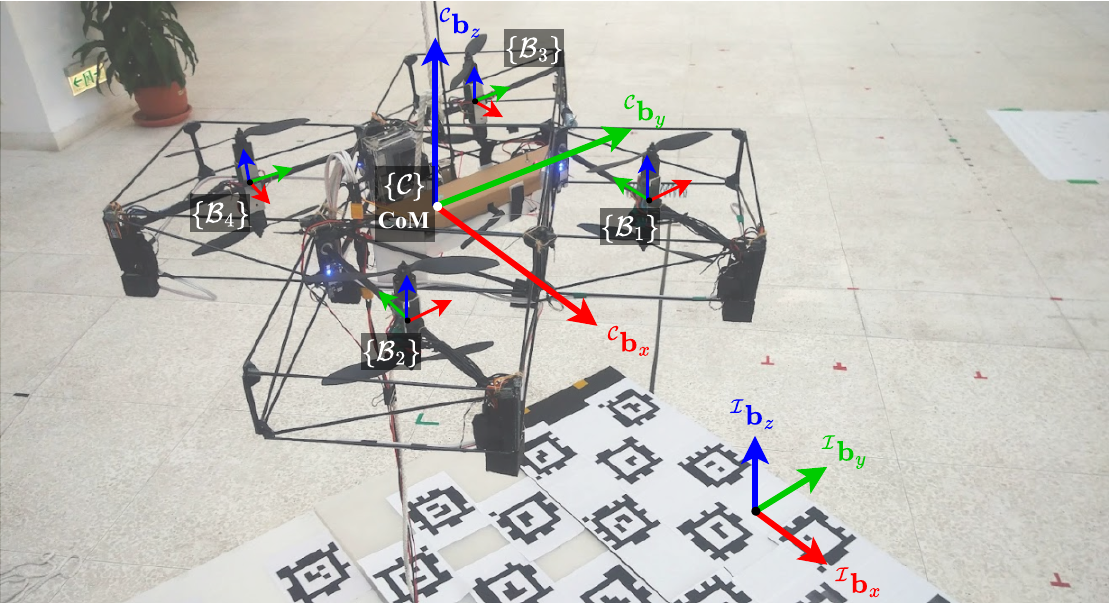}
    \caption{The prototype of the proposed A4-\textit{Inc} team UAV, which is a 4-agent system with one additional navigator module at the center. The agent 1 and agent 2 are placed in inconsistent orientation. \label{fig:team_prototype}}
\end{figure}

In this paper, a thrust vectoring modular drone (TVMD) is proposed (\figref{fig:team_prototype}),  which is a team system composed of thrust-vectoring agents with coaxial rotors, where the gyroscopic moment can be reduced thanks to the cancelled angular momentum. 
The coaxial rotor design prevents the team inertia growing rapidly with the number of agents while remaining high power density \cite{qinGeminiCompactEfficient2020}.
The frame-bounded structure is also safer as a collision occurs\footnote{These agents are designed to be capable of flying standalone with the thrust vectoring control (TVC) mechanism, but this is beyond the scope of this article.}.
The over-actuated nature provides redundancy in case of failures, 
and the degree of redundancy can be increased by adding more modules in a team according to the required level of safety.
With the modular design, TVMD has the ability to be reconfigured according to specific task requirements, and thus, the efficiency or performance can be optimized according to the payload weights, destination distance, weather conditions, etc. 
\figref{fig:team_prototype} shows the prototype of the proposed TVMD in a 4-agent team.
Let $\{\FVi{I}{b}{x}, \FVi{I}{b}{y}, \FVi{I}{b}{z}\}$, $\{\FVi{C}{b}{x}, \FVi{C}{b}{y}, \FVi{C}{b}{z}\}$, and $\{\,^{\mathcal{B}_i}\mathbf{b}_x, \,^{\mathcal{B}_i}\mathbf{b}_y, \,^{\mathcal{B}_i}\mathbf{b}_z\}$ be the basis vectors of the inertial frame $\{\mathcal{I}\}$, the team vehicle frame $\{\mathcal{C}\}$ with its origin defined at the center of mass (CoM) of the team, and the body frame of the $i$-th agent module $\{\mathcal{B}_i\}$, respectively, where $i \in {1..n}$ and $n$ is the total number of agents.
The orientation of agents in a team is assumed to have discrete angles. 
A team configuration is defined as 
\begin{equation}
    \Xi = \left\{(\mathbf{p}_i, \psi_i)\middle| 
    \begin{array}{l}
        \mathbf{p}_i \in \mathbb{R}^3,  \\
        \psi_i=\left\{m {\pi}/{2} | m \in 0, 1, 2,...\right\} 
    \end{array}
    \right\},
    \label{eq:team_conf}
\end{equation} 
where $\mathbf{p}_i$ and $\psi_i$ are the position and the z-axis orientation of agent $i$ (the $\{\mathcal{B}_i\}$ frame) relative to the team frame $\{\mathcal{C}\}$, respectively. 
Rotations between $\{\mathcal{C}\}$ and $\{\mathcal{B}_i\}$ are described by $\,^\mathcal{C}\mathbf{R}_{\mathcal{B}_i}=\mathbf{R}_z(\psi_i)$, where $\mathbf{R}_z$ is the basic rotation along the z-axis.
If all the agents are in the same directions, i.e., $m$ is even for all $i$, the team is called a \textit{consistent} orientation configuration (\textit{Con}), and otherwise, it is called an \textit{inconsistent} orientation configuration (\textit{Inc}).

\begin{figure}[btp]
    \centering
    \includegraphics[width=1.0\linewidth, trim={{0cm} {0cm} {1cm} {0.5cm}}, clip]{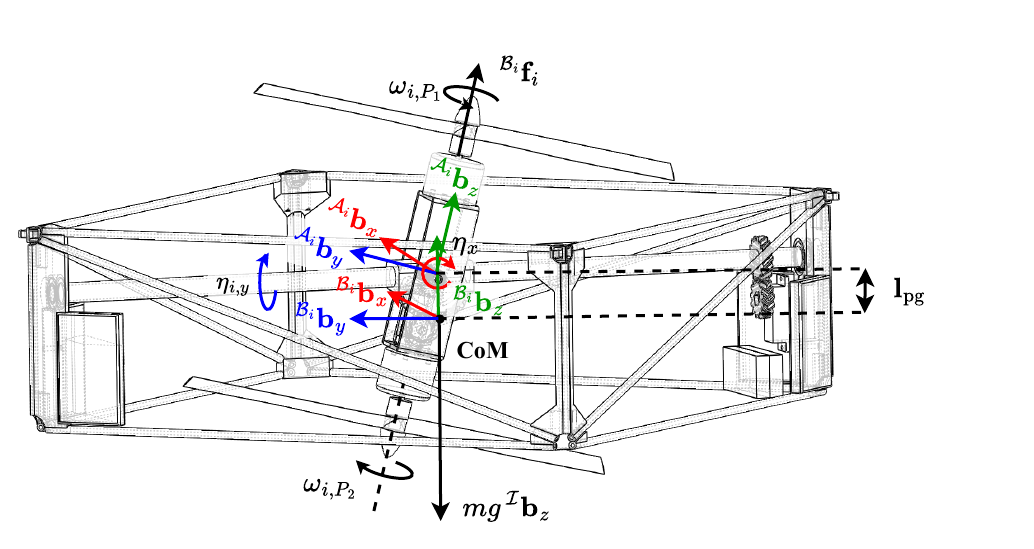}
    \caption{The design of a single agent. The gimbal design leads to inherent distinct angle limits on $\eta_{i,x}$ and $\eta_{i,y}$, which results in different orientations in a team system and forms complicated attainable force space.\label{fig:single_free_body}}
\end{figure}

The design and the coordinate systems of a single agent system are depicted in \figref{fig:single_free_body}.
The actuator frame $\{\mathcal{A}_i\}$ of agent $i$ is driven by x and y-axis servo motors, and the angles of rotations are denoted by $\eta_{i,x}$ and $\eta_{i,y}$, respectively.
The pointing direction of the thrust vector of an agent, i.e., $\,^{\mathcal{A}_i}\mathbf{b}_z$ expressed in $\{\mathcal{B}_i\}$, is described by a reduced attitude 
\begin{equation}
    \hat{\mathbf{a}}_i = \BiRAi 
    \begin{bmatrix}
        0 \\ 0 \\ 1
    \end{bmatrix}
    = \begin{bmatrix}
        \cos\eta_{i,x} \sin\eta_{i,y} \\
        -\sin\eta_{i,x}           \\
        \cos\eta_{i,x} \cos\eta_{i,y}
    \end{bmatrix} \in \mathsf{S}^2,
    \label{eq:model:gimbal_reduced_attitude}
\end{equation}
where $\mathsf{S}^2$ is the unit sphere. 
\subsection{System Dynamics}



The team vehicle is assumed to be rigidly connected, and the team dynamics are
\begin{align}
    \dFVB{}{x}{C} &= \FVB{}{R}{C} \FVB{}{v}{C}, \quad
    \dFVB{}{R}{C} = \FVB{}{R}{C} [\FVB{}{\Omega}{C}]_\times, 
    \label{eq:model:team:positions}\\
    G_\mathcal{C} \dFVB{}{\xi}{C} &= C(\FVB{}{\xi}{C}) \FVB{}{\xi}{C} + \,^\mathcal{C}\mathbf{u} + \,^\mathcal{C}\mathbf{g}, 
    \label{eq:model:team:rotations} \\
    C(\xi_\mathcal{C}) &= \begin{bmatrix}
        [\mathbf{J}_\mathcal{C}\Omega_\mathcal{C}]_\times & \mathbf{0}_3 \\
        \mathbf{0}_3 & -[m_\mathcal{C}\Omega_\mathcal{C}]_\times
    \end{bmatrix}, \nonumber\\
    G_\mathcal{C} &= \begin{bmatrix}
        \mathbf{J}_\mathcal{C} & \mathbf{0}_3 \\
        \mathbf{0}_3 & m_\mathcal{C}\mathbf{I}_3
    \end{bmatrix}, \quad
    \,^\mathcal{C}\mathbf{g} = \begin{bmatrix}
        \mathbf{0}_3 \\ m_\mathcal{C}g \mathbf{R}_\mathcal{C}^\top \mathbf{e}_3
    \end{bmatrix}, \nonumber
\end{align}
where $\mathbf{x}_\mathcal{C} \in \mathbb{R}^3$ and $\mathbf{v}_\mathcal{C} \in \mathbb{R}^3$ is the position and velocity of the team vehicle CoM expressed in the inertial frame, respectively.
$\xi_\mathcal{C} = \begin{bmatrix} \Omega_\mathcal{C}^\top \quad \mathbf{v}_\mathcal{C}\top \end{bmatrix}^\top$ is the twist of the team vehicle.
$[\cdot]_\times$ denotes the skew-symmetric operator. 
$G_\mathcal{C}$ is the spatial inertia matrix of the team system, where $m_\mathcal{C}=m_0 + \sum_i^n m_i$ and $\mathbf{J}_\mathcal{C} = \mathbf{J}_0 - \sum_i^n m_i [\mathbf{p}_i]_\times [\mathbf{p}_i]_\times$ are the team mass and the moment of inertia, respectively, 
with the subfix $0$ denoting for the navigator module.
$C(\FVB{}{\xi}{C})$ is a skew-symmetric matrix containing fictitious forces, 
and $\FVB{B}{g}{} \in \mathbb{R}^6$ is the gravity wrench.
%
To reduce the complexity in control allocation, only thrust vectors from agents are considered, and the effect of gyroscopic moment and propeller drag moments are neglected. 
The wrench control input is modeled by
\begin{align}
    \,^\mathcal{C}\mathbf{u} 
    = \begin{bmatrix}
        \,^\mathcal{C}\mathbf{u}_\tau \\
        \,^\mathcal{C}\mathbf{u}_f 
    \end{bmatrix}
    = \sum_i^n \begin{bmatrix}
        \mathbf{p}_i \times (\,^\mathcal{C}\mathbf{R}_{\mathcal{B}_i} T_{f_i} \hat{\mathbf{a}}_i) \\
        \,^\mathcal{C}\mathbf{R}_{\mathcal{B}_i} T_{f_i}\hat{\mathbf{a}}_i
    \end{bmatrix} 
    = \mathbf{M}(\Xi) \mathbf{f},
    \label{eq:model:team:effectiveness_model}
\end{align}
where $\mathbf{M}(\Xi) \in \mathbb{R}^{6\times 3n}$ is the effectiveness matrix and $\mathbf{f}=[\mathbf{f}_1^\top .. \mathbf{f}_n^\top]^\top$ is the control force vector. 
$\mathbf{f}_i  \in \mathbb{R}^{3}$ is the thrust vector of agent $i$, which is modeled by the Momentum theory and the reduced attitude of $\{\mathcal{A}_i\}$ described in \eqref{eq:model:gimbal_reduced_attitude}:
\begin{equation}
    \mathbf{f}_i = {T_f}_i \hat{\mathbf{a}}_i, \quad
    {T_f}_i = \rho d^4 c_l \omega_i^2,
    \label{eq:model:thrust_force}
\end{equation}
where $T_{f_i}$ is the thrust generated by agent $i$, $\rho$ is the air density, $d$ is the propeller diameter, $c_l$ is the lift coefficient, and $\omega_i = \omega_{i, P_1} = \omega_{i, P_2}$ is the spinning speed of the propeller.
The actual command of the team vehicle is $(\eta_{i,x}, \eta_{i,y}, \omega_i)$ for all agents, which is constrained by 
\begin{equation}
    \begin{pmatrix}
        \eta_{i,x} \\ \eta_{i,y} \\ \omega_i
    \end{pmatrix}
    \in \mathcal{E}_f = \left\{
    \begin{pmatrix}
        \eta_{i,x} \\ \eta_{i,y} \\ \omega_i
    \end{pmatrix}
    \middle|
    \begin{array}{l}
        \eta_{i,x} \in [-\sigma_x, \sigma_x] \\ 
        \eta_{i,y} \in [-\sigma_y, \sigma_y]\\ 
        \omega_i \in (0, \sigma_{\omega \max}]
    \end{array}
    \right\},
    \label{eq:model:control_space}
\end{equation}
where $\mathcal{E}_f$ is the admissible control space of the team.

\section{Full-Pose Tracking Control}

\begin{figure}[btp]
    \centering
    \includegraphics[width=1.0\linewidth, trim={0 0 0 0}, clip]{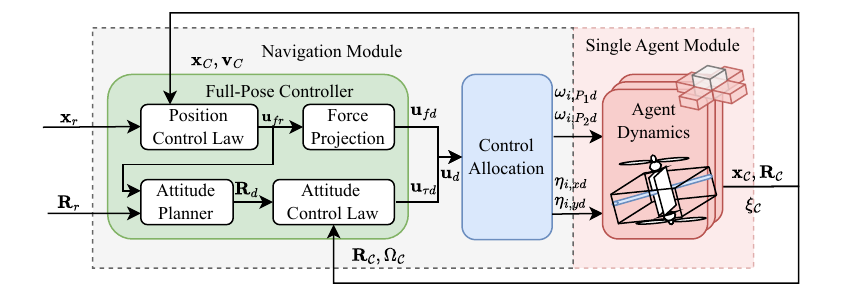}
    \caption{The control block diagram of the team system.}
    \label{fig:ctrl:block_diagram}
\end{figure}

To design a control strategy for general team configurations, a cascaded control architecture is adopted, which features the separation of tracking performance design and the team configuration \cite{invernizziComparisonControlMethods2021}.
However, the team configuration defines the attainable controls that a tracking controller can utilize, which motivates us to approximate the attainable force set (AFS) $\tilde{\mathcal{U}}_f (\Xi)$.
\figref{fig:ctrl:block_diagram} illustrates the overall control block diagram.
The full-pose tracking controller aims to design a virtual control, $\mathbf{u}_d$, with the consideration of internal state constraints for a variety of team configurations.
Then, the control allocation (CA) is followed in charge of the fulfillment of $\mathbf{u}_d$ by calculating the actual control inputs $(\eta_{i,x}, \eta_{i,y}, \omega_i)$.
%

%


\subsection{Attainable Force Space (AFS) Approximation}

\begin{figure}[btp]
    \centering
    \includegraphics[width=0.85\linewidth, trim={{2cm} {2.7cm} {2.1cm} 0}, clip]{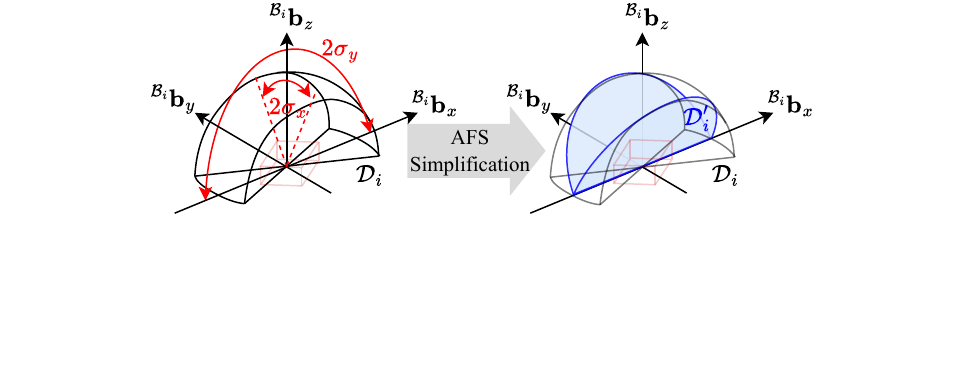}
    \caption{The spherical slices simplification of a single agent AFS. With $\sigma_x=\pi/6$, $\sigma_y=\pi/2$, agent $i$'s AFS, $\mathcal{D}_i$, is non-convex as drawn in the left figure, which can be approximated by a spherical slice $\mathcal{D}_i'$.}
    \label{fig:model:single_AFS_slices}
\end{figure}

\begin{figure}[btp] 
    \centering
    \subfloat[Consistent orientation agents, i.e., $n_x=2$, $n_y=0$. \label{fig:model:team_AFS_con}]{%
    \includegraphics[width=0.9\linewidth, trim={{2cm} {1cm} {1.5cm} 0}, clip]{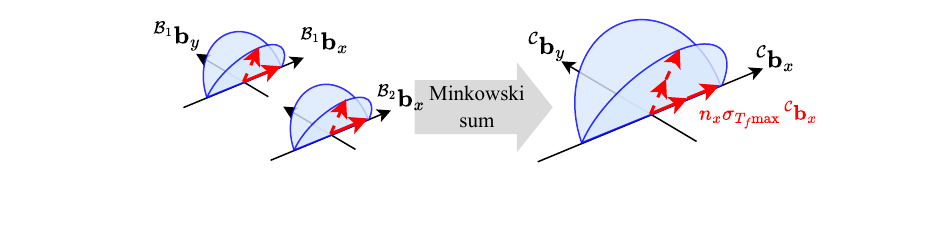}}
\vfill
    \subfloat[Inconsistent orientation agents, i.e., $n_x=1$, $n_y=1$. \label{fig:model:team_AFS_inc}]{%
    \includegraphics[width=0.9\linewidth, trim={{2cm} {1cm} {1.5cm} 0}, clip]{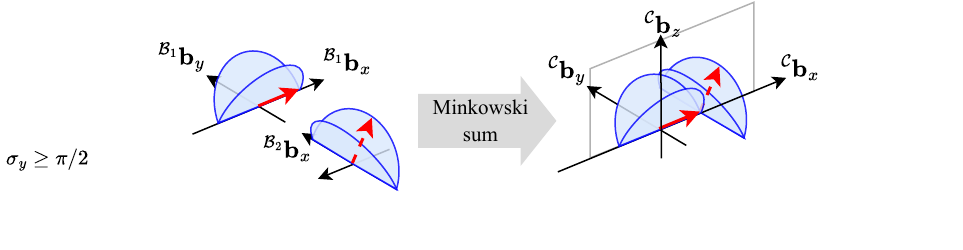}}
\caption{Examples of Minkowski sum of AFSs with $n=2$, $\sigma_y = \pi/2$, and $\sigma_x < \pi/2$. For the special case that either $\sigma_x = \pi/2$ or $\sigma_y = \pi/2$, the bottom of the concated AFS is not a point but a line for consistent orientation agents or a plane for inconsistent orientation agents.}
\label{fig:model:team_AFS}
\end{figure}

The AFS of the team vehicle 
$
    {\mathcal{U}_f} = \left\{
        \,^\mathcal{C}\mathbf{u}_f \middle| (\eta_{i,x}, \eta_{i,y}, \omega_i) \in \mathcal{E}_f
    \right\}
$
is defined by the set that the force control can reach according to \eqref{eq:model:team:effectiveness_model}, \eqref{eq:model:thrust_force}, and \eqref{eq:model:control_space}.
However, the actual form of $\mathcal{U}_f$ is hard to obtain since \eqref{eq:model:thrust_force} is nonlinear and the constraints $\mathcal{E}_f$ lead to non-convex AFS.
And therefore, spherical slices drawn in \figref{fig:model:single_AFS_slices} are used to approximate the AFS of a single agent.
And then the concatenation of AFSs can be calculated by Minkowski sum \cite{orrHighEfficiencyThrustVector2014} since the spherical slices are convex.
\figref{fig:model:team_AFS} shows examples for consistent and inconsistent agent orientation in a 2-agent team.
Thanks to the constraints on agent orientations $\psi_i$ in \eqref{eq:team_conf}, the Minkowski sum, i.e., the approximated AFS, can be approximated geometrically by an elliptic cone which is given by
\begin{align}
    \tilde{\mathcal{U}}_f(\Xi, s)
     & = \left\{\mathbf{u}_f \in \mathbb{R}^3 \middle| 
     \begin{array}{l}
        \mathbf{u}_f^\top \mathbf{C}(\mathbf{u}_{f,z}) \mathbf{u}_f \leq 1, \\
        \|\mathbf{u}_{f}\|_2^2 < n^2 \sigma_{T_f}^2    
     \end{array}
     \right\}, 
    \label{eq:swarm:attainable_space_apprx} \\
    \mathbf{C}(\mathbf{u}_{f,z}) &= \text{diag}\left({1}/{c_x^2(\mathbf{u}_{f,z}, s)}, {1}/{c_y^2(\mathbf{u}_{f,z}, s)}, 0\right), \nonumber
\end{align}
where the maximum thrust $\sigma_{T_f} = \rho d^4 c_l \sigma_\omega^2$, and the semi-axes are 
\begin{align}
    c_x(z, s)          & = g(n_x, s\sigma_y, z) + g(n_y, s\sigma_x, z),
    \label{eq:model:team:attainable_space_approx_semimajor_axis_x} \\
    c_y(z, s)          & = g(n_x, s\sigma_x, z) + g(n_y, s\sigma_y, z), 
\label{eq:model:team:attainable_space_approx_semimajor_axis_y} \\
    g(n_k, \sigma, z) & = \left\{
    \begin{array}{ll}
        n_k \sigma_{T_f}             & \text{if}\,\sigma \geq \pi/2 \\
        (n_k/n)(1-\delta(n_k))|z|\tan\sigma & \text{otherwise}
    \end{array}
    \right. \nonumber
\end{align}
where $n_x$ is the number of agents with their orientation the same with $\{\mathcal{C}\}$, i.e., $\psi_i = \{m\pi/2\}$, and $m$ is even, and then $n_y = n-n_x$.
$g(n_k, \sigma, z)$ is the portion contributed by $n_k$ agents with maximum deflection limit $\sigma$ at height $z$, $k \in \{x, y\}$.
$s \in (0, 1]$ is a relaxation which narrows the approximated AFS $\tilde{\mathcal{U}}_f$ and avoids controls appearing near the real AFS boundary. The effect of $s$ will be delivered in \secref{sec:sim:tilting:relax}.
$\delta(\cdot)$ is the delta function which is all zero except for the origin.
$\tilde{\mathcal{U}}_f$ is a function of $z$ since the AFS of a single agent is also a function of $z$.
Online re-calculation of $\mathbf{C}(\mathbf{u}_{f,z})$ is invoked in the attitude planner and the force projection in each control cycle.
%

\subsection{Attitude Planner\label{sec:ctrl:team:attitude_planner}}

Since the approximated attainable force space $\tilde{\mathcal{U}}_f$ of a team does not extend to every direction in $\mathbb{R}^3$, the reference attitude $\mathbf{R}_r = [\mathbf{R}_{xr} \,\,\, \mathbf{R}_{yr} \,\,\, \mathbf{R}_{zr}]$ may be infeasible to fulfill the required forces $\mathbf{u}_{fr}$ to track the reference trajectory $\mathbf{x}_r$. 
Thus, an attitude planner is used to design a temporary attitude target,
\begin{equation}
    \mathbf{R}_d = \begin{cases}
        \mathbf{R}_r,   & \text{if } \mathbf{R}_r^\top \mathbf{f}_r \in \tilde{\mathcal{U}}_f, \\
        \begin{bmatrix}
            \mathbf{b}_{xd} & \mathbf{b}_{yd} & \mathbf{b}_{zd}
        \end{bmatrix}, & \text{otherwise}
    \end{cases},
    \label{eq:ctrl:team:fullpose:attitude_planner}
\end{equation}
where $\mathbf{f}_r = t_{T_f} \mathbf{R}_\mathcal{C} \mathbf{u}_{fr}$ is the required force expressed in $\{\mathcal{I}\}$.
The $t_{T_f}$, given by 
\begin{equation}
    t_{T_f} = \min \left( 1, n \sigma_{T_f} / \| \mathbf{u}_{fr}\|_2 \right),
    \label{eq:ctrl:team:fullpose:force_scaling}
\end{equation}
scales $\mathbf{u}_{fr}$ to the maximum thrust $n \sigma_{T_f}$ if it exceeds the limit, which ensures the solution to \eqref{eq:ctrl:team:fullpose:attitude_planner} always exists.
The soluton to $\mathbf{R}_d$ when $\mathbf{R}_r^\top \mathbf{f}_r \notin \tilde{\mathcal{U}}_f$ is formulated as an optimization problem similar to the one in \cite{franchiFullPoseTrackingControl2018}, but the elliptic conic $\tilde{\mathcal{U}}_f$ is used,
\begin{equation}
    \theta_\star = \underset{\theta}{\arg\min} 
    \left\{ -{\mathbf{b}}_{zr}^\top {\mathbf{b}}_{z\parallel}'(\theta)
    \middle| 
    \begin{array}{ll}
        \mathbf{R}'^\top \mathbf{f}_{r} \in \tilde{\mathcal{U}}_f, \\
        \mathbf{R}' \in \mathsf{SO}(3)
    \end{array}
    \right\},
    \label{eq:team:fullpose:simplified_opt_problem}
\end{equation}
where $\mathbf{b}_{z\parallel}'(\theta)$ is a vector rotated by $\theta$ from $\mathbf{b}_{zr}$ on the plane formed by $\mathbf{b}_{zr}$ and $\mathbf{f}_{r}$ \footnote{For the special case that $\mathbf{b}_{zr}$ is almost at the opposite direction of $\mathbf{f}_r$, their coplane will be hard to define, and then a arbitray vector on x-y plane can be used according to preferences of vehicle designs.}, which can be calculated using Rodrigues' rotation formula.
    Since $\frac{d}{d\theta}(-{\mathbf{b}}_{zr}^\top {\mathbf{b}}_{z\parallel}'(\theta))=\sin\theta$, $-{\mathbf{b}}_{zr}^\top {\mathbf{b}}_{z\parallel}'(\theta)$ is strictly monotonic along $\theta$ in $[0, \pi)$.
    The solution to \eqref{eq:team:fullpose:simplified_opt_problem} always exists.
\eqref{eq:team:fullpose:simplified_opt_problem} is solved using a bisection algorithm \cite{franchiFullPoseTrackingControl2018}, but instead of using $\theta_{\max}=\arcsin(\|\mathbf{k}\|)$, to extend all possible attitudes in $[0, 2\pi)$, $\theta_{\max} = \arccos(\mathbf{b}_{zr}^\top \mathbf{f}_r / \| \mathbf{f}_r \|_2)$ is used.
Finally, $\mathbf{b}_{zd} = {\mathbf{b}}_{z\parallel}'(\theta_\star)$, $\mathbf{b}_{yd}=\frac{\mathbf{b}_{zd} \times \mathbf{b}_{1r}}{\|\mathbf{b}_{zd} \times \mathbf{b}_{1r}\|}$, and $\mathbf{b}_{xd}=\mathbf{b}_{yd} \times \mathbf{b}_{zd}$.
And then $\mathbf{R}_d$ is guaranteed to be capable of fulfilling $\mathbf{f}_{r}$ under the given team configuration, or $t_{T_f} \mathbf{R}_d^\top \mathbf{R}_\mathcal{C} \mathbf{u}_{fr} \in \tilde{\mathcal{U}}_f$. 
Note that as discussed in \eqref{eq:swarm:attainable_space_apprx}, the approximated AFS can be tuned such that the required force will not be too close to the boundary of the actual AFS, which eliminates the possibility of the vehicle entering the infeasible attitude region.
Namely, the force will not be frequently saturated by the force projection, which is a trade-off between the attitude tracking performance and robustness.




\subsection{The Control Law}
\begin{figure}[btp] 
    \centering
    \subfloat[Scaling by $t_{T_f}$. \label{fig:ctrl:mag_projection}]{%
    \includegraphics[width=0.36\linewidth, trim={0 0 {0.8cm} 0}, clip]{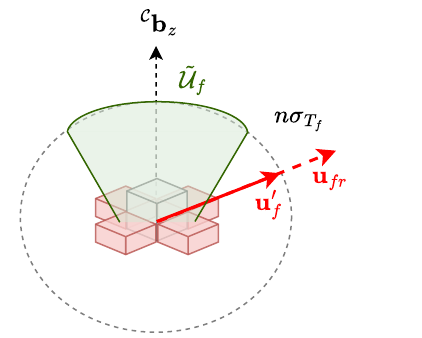}}
\hfill
    \subfloat[Projection by $t_\eta$. \label{fig:ctrl:dir_projection}]{%
    \includegraphics[width=0.64\linewidth, trim={0 {0.4cm} {0.9cm} 0}, clip]{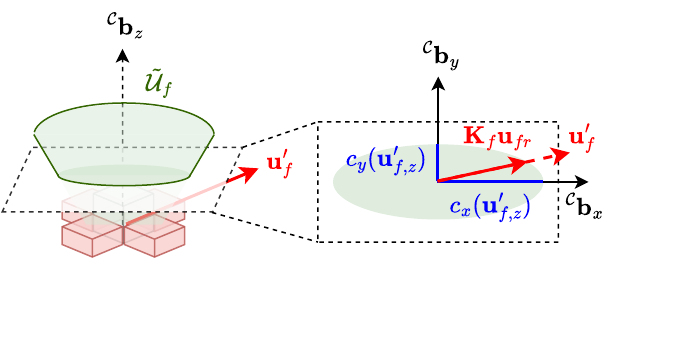}}
\caption{Illustration of force projection $\mathbf{K}_f$. The force projection firstly scales a given force $\mathbf{u}_{fr}$ by $t_{T_f}$, and then projects it to ellipse at the height of $\mathbf{u}_{fr}'$.}
\label{fig:ctrl:force_projection}
\end{figure}

The control law aims to enforce $(\mathbf{x}_\mathcal{C}, \mathbf{R}_\mathcal{C}, \xi_\mathcal{C}) \to (\mathbf{x}_r, \mathbf{R}_d, \xi_d)$, where $\mathbf{x}_r$ is the reference trajectory.
The position tracking error is $\mathbf{e}_\mathbf{x} = \FVB{}{x}{C} - \mathbf{x}_r$.
By defining the attitude error function 
$
\Psi(\FVB{}{R}{C}, \mathbf{R}_d) = \frac{1}{2} \text{tr} \left(\mathbf{K}_\mathbf{R}(\mathbf{I} - \mathbf{R}_d^\top \FVB{}{R}{C})\right)
$, $\mathbf{K}_\mathbf{R} \succ 0$
the attitude error is 
$
\mathbf{e}_R = \frac{1}{2}
(\mathbf{R}_d^\top \FVB{}{R}{C} - \FVB{}{R}{C}^\top \mathbf{R}_d)^\vee  \in \mathbb{R}^3
$, where $(\cdot)^\vee$ is the inverse mapping of $[\cdot]_\times$.
The twist error is 
\begin{align*}
    \mathbf{e}_\xi & = \FVB{}{\xi}{C} - \xi_d
    = \begin{bmatrix}
          \mathbf{e}_\Omega \\
          \mathbf{e}_v
      \end{bmatrix}
    = \begin{bmatrix}
          \FVB{}{\Omega}{C} \\ \FVB{}{v}{C}
      \end{bmatrix}
    - \begin{bmatrix}
          \FVB{}{R}{C}^\top \mathbf{R}_d \Omega_d \\
          \FVB{}{R}{C}^\top \dot{\mathbf{x}}_r
      \end{bmatrix}.
\end{align*}
%
Define the Lyapunov candidate for the tracking error,
$
\Phi = \frac{1}{2} \mathbf{e}_\mathbf{x}^\top \mathbf{K}_\mathbf{x} \mathbf{e}_\mathbf{x} + \Psi(\FVB{}{R}{C}, \mathbf{R}_d) 
$, where $\mathbf{K}_\mathbf{x}$ is a diagonal weight matrix.
%
Using similar strategies in \cite{nguyenNovelRoboticPlatform2018} in additin to a force projection $\mathbf{K}_f$, the control law is given by 
\begin{align}
    \mathbf{u}_d & = \begin{bmatrix}
        \mathbf{I}_3 & \mathbf{0}_3 \\
        \mathbf{0}_3 & \mathbf{K}_f
     \end{bmatrix} \mathbf{u}_r, \quad 
     \mathbf{K}_f = \text{diag}(t_\eta t_{T_f}, t_\eta t_{T_f}, t_{T_f})
    \label{eq:ctrl:team:lay:control_law} \\
    \mathbf{u}_r & = G_C \dot{\xi}_d - C(\xi) \xi_d - \mathbf{K}_\xi \mathbf{e}_\xi - \nabla\Phi - \,^\mathcal{B}\mathbf{g}, \nonumber
\end{align}
provided that $\nabla\Phi = \begin{bmatrix}(\mathbf{K}_\mathbf{R} \mathbf{e}_\mathbf{R})^\top & (\mathbf{R}_\mathcal{C}^\top \mathbf{K}_\mathbf{x} \mathbf{e}_\mathbf{x})^\top\end{bmatrix}^\top$, and $\mathbf{K}_\xi \succ 0$. 
$\mathbf{u}_r = \begin{bmatrix} \mathbf{u}_{\tau r}^\top & \mathbf{u}_{fr}^\top \end{bmatrix}^\top \in \mathbb{R}^6$ is the unconstrained wrench control.
To ensure $\mathbf{u}_d \in \tilde{\mathcal{U}}_f$ when the current attitude is infeasible to fulfill $\mathbf{u}_{fr}$, the force projection $\mathbf{K}_f$ firstly scales $\mathbf{u}_{fr}$ by $t_{T_f}$  (see \figref{fig:ctrl:mag_projection}) if $\mathbf{u}_{fr}$ is out of the maximum available thrust, 
and then a projection about x and y-axis at the height of $\mathbf{u}_{f}'$ (see \figref{fig:ctrl:dir_projection}) is followed, which is done by 
\begin{align}
    t_\eta &=
    \min\left(
        1, 
        \sqrt{ 1 / \mathbf{u}_f'^\top \mathbf{C}(\mathbf{u}_{f,z}') \mathbf{u}_f' }
    \right), \quad
    \mathbf{u}_{f}' = t_{T_f} \mathbf{u}_{f},
    \label{eq:fullpose:dir_sat}
\end{align}
where $\mathbf{C}(\mathbf{u}_{f,z}')$ is defined in \eqref{eq:swarm:attainable_space_apprx} with $s=1$ such that the full AFS can be utilized.
The projections are illustrated in \figref{fig:ctrl:force_projection}.
Note that $t_\eta = 1$ as $\mathbf{u}_{fz}' \in \tilde{\mathcal{U}}_f$, which can be checked by substituting \eqref{eq:swarm:attainable_space_apprx} into \eqref{eq:fullpose:dir_sat}.
The force projection is crucial for the control allocation to ensure the solution exists.

The convergence of the tracking error is analyzed by considering the error dynamics. By substituting \eqref{eq:ctrl:team:lay:control_law} into \eqref{eq:model:team:rotations}, one obtaines
\begin{align}
    G_\mathcal{C} \dot{\mathbf{e}}_\xi = C(\xi_\mathcal{C})\mathbf{e}_\xi - \mathbf{K}_\xi \mathbf{e}_\xi - \nabla\Phi
    + 
    \begin{bmatrix}
    \mathbf{0}_3 & \mathbf{0}_3 \\
    \mathbf{0}_3 & \mathbf{K}_f - \mathbf{I}_3
    \end{bmatrix}
    \mathbf{u}_r.
    \label{eq:ctrl:team:lya:closed_loop_dyn}
\end{align}
The only difference with the one in \cite{nguyenNovelRoboticPlatform2018} is the additional $\mathbf{K}_f - \mathbf{I}_3$ in translational part.
Since the upper-right block of $C(\xi_\mathcal{C})$ is $\mathbf{0}_3$, the asymptotic convergence of attitude does not affected by the position and velocity errors.
Once the attitude error converges, i.e., $\mathbf{R}_\mathcal{C} \to \mathbf{R}_d$, as discussed in \secref{sec:ctrl:team:attitude_planner}, $t_\eta \to 1$, and then the fulfillment of $t_{T_f} \mathbf{u}_{fr}$ is guaranteed.
Assuming that the reference trajectory $\mathbf{x}_r$ is properly designed such that the magnitude of the nominal force to track the reference trajectory $\mathbf{x}_r$ is bounded by $n \sigma_{T_f}$, $t_{T_f}$ will then become 1 after the position error is small enough. 
And then, the rest of \eqref{eq:ctrl:team:lya:closed_loop_dyn} degrade to the one in \cite{nguyenNovelRoboticPlatform2018}, which is proven to be global asymptotic stable except for the unstable equilibrium points. 

\subsection{Control Allocation}
The under-determined inversion of \eqref{eq:model:team:effectiveness_model} and \eqref{eq:model:thrust_force} can be solved using Redistributed pseudo-inverse \cite{jinModifiedPseudoinverseRedistribution2005} or constrained nonlinear optimization methods \cite{suDownwashawareControlAllocation2022a}, however, it is out of the scope of this paper. It is assumed that $\mathbf{f}_d=[\mathbf{f}_{d1}^\top..\mathbf{f}_{dn}^\top]^\top$ is feasible as the control $\mathbf{u}_d = \mathbf{M}(\Xi) \mathbf{f}_d \in \mathcal{U}_f$ and the constraints in \eqref{eq:model:control_space} are guaranteed by the proposed controller, and then the actual controls are
\begin{align}
    {T_f}_{i}  &= \|\mathbf{f}_{di}\|_2,          \nonumber\\
    \eta_{i,x} &= \arcsin\left(
    {-\mathbf{f}_{di,y}}/{{T_f}_{i}}\right), \nonumber\\
    \eta_{i,y} &= \arctan2\left(
    {\mathbf{f}_{di,x}}/{\text{sgn}(\cos\eta_{i,x})},
    {\mathbf{f}_{di,z}}/{\text{sgn}(\cos\eta_{i,x})}\right), \nonumber
\end{align}
where $\text{sgn}(\cdot)$ is the sign function. 
The singularity occurs at $\eta_{i,x} = \pm\pi/2$, which is avoided inherently by the constraints of gimbal mechanisms as in \figref{fig:single_free_body}.

\section{Simulations\label{sec:simulations}}

\begin{figure}[btp] 
    \centering
    \includegraphics[width=1.0\linewidth, trim={0 0 0 0}, clip]{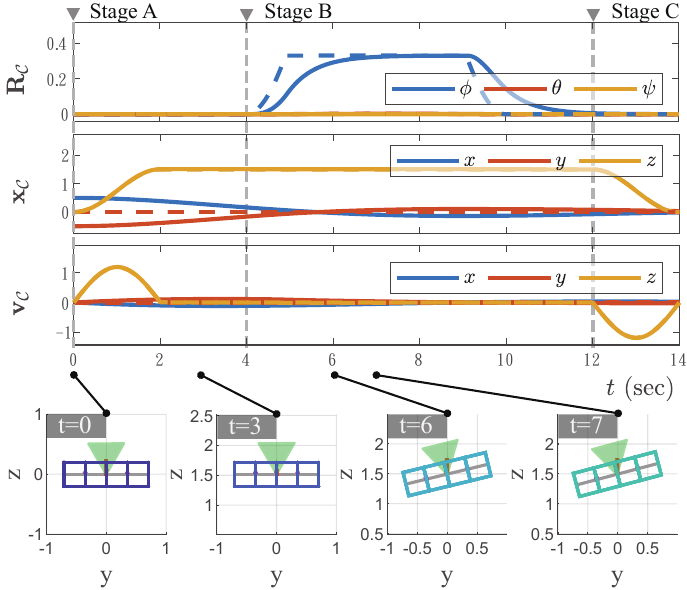}
\caption{The state profile in the simulation, where the dashed lines represent for the desird states, including $\mathbf{x}_r$, $\dot{\mathbf{x}}_r$, and the planned attitude $\mathbf{R}_d$. The attitude $\mathbf{R}_\mathcal{C}$ is expressed in Euler angles for easier understanding. The green cones indicate the approximated AFS using \eqref{eq:swarm:attainable_space_apprx}. \label{fig:sim:tilting:state}}
\end{figure}

%
The simulations are held in \textit{Matlab} \textit{Simulink}, the dynamics are continuous and the controller is digital with sampling time $\Delta t=0.01$ second.
An A4-\textit{Inc} team UAV with dynamics \eqref{eq:model:team:positions} and \eqref{eq:model:team:rotations} is used. The internal dynamics i.e., servos and propellers are assumed to respond immediately and internal wrenches are not considered. 
The control constraints are $\sigma_{x}=\pi/6$, $\sigma_{y}=\pi/4$, and $\sigma_{T_f}=2 m_i g$.
The controller gains are 
$\mathbf{K}_\mathbf{x} = \text{diag}(0.4, 0.4, 1)$, 
$\mathbf{K}_\mathbf{R}=\text{diag}(12, 12, 1)$, 
$\mathbf{K}_\xi=\text{diag}(8, 8, 1.5, 0.8, 0.8, 2)$.
The initial position is $\mathbf{x}_0=[0.5 \,\,\, -0.5 \,\,\, 0]^\top$.
There are three stages in the reference trajectory: 
A) Ascending, the vehicle tracks $\mathbf{x}_{r,z}(t) = 0.75(1 - \cos(t \pi / 2))$ in the first 2 seconds to reach 1.5 meters height, and then hovers at $[ 0\,\,\, 0 \,\,\, 1.5]^\top$. 
B) Tilting, the vehicle tracks a roll angle $\phi_r(t) = (\pi/6)(1 + \cos((t - 7)\pi/3))$ from $t=4$ to $t=10$. Note that the maximum roll angle $\pi/3$ is out of feasible attitude space due to the limited gimbal angle.
C) Descending, the vehicle follows the same path in ascending stage but in a reverted direction to return to the origin.
For the reference positions, second-order commands are used, i.e., $\mathbf{x}_r, \dot{\mathbf{x}}_r, \ddot{\mathbf{x}}_r$ are provided, as for the rotational part, only the attitude command is given, i.e., $\phi_r(t)$ and $\Omega_d=\mathbf{0}$, and a planner relaxation $s=0.5$ is used.
The reference trajectory is drawn in dashed lines in \figref{fig:sim:tilting:state}.

\begin{figure}[btp] 
    \centering
    \includegraphics[width=0.9\linewidth, trim={0 0 0 0}, clip]{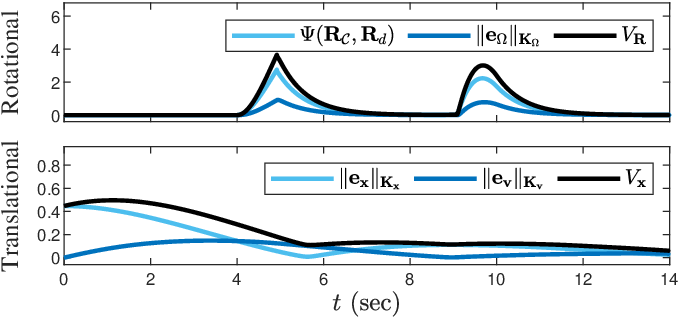}
\caption{The Lyapunov candidates profile in the simulation, where $V_\mathbf{R} = \Psi(\mathbf{R}_\mathcal{C}, \mathbf{R}_d) + \Vert\mathbf{e}_\Omega\Vert_{\mathbf{K}_\Omega}$ and $V_\mathbf{x} = \Vert\mathbf{e}_\mathbf{x}\Vert_{\mathbf{K}_\mathbf{x}} + \Vert\mathbf{e}_\mathbf{v}\Vert_{\mathbf{K}_\mathbf{v}}$. \label{fig:sim:tilting:lyapunov}}
\end{figure}

Due to the constraints of the approximated AFS $\tilde{\mathcal{U}}_f$, the attitude target $\mathbf{R}_d$ is dominated by $\theta_\star$ in \eqref{eq:team:fullpose:simplified_opt_problem}, which limits the maximum tilting angle of $\phi_d$ to around $\pi/12$.
Thus, the vehicle maintains the stability in positional responses by ensuring the fulfillment of forces by avoiding entering infeasible attitude regions.
The force projection \eqref{eq:ctrl:team:lay:control_law} plays an crucial role to ensure the control allocation produce predictible results.
From \figref{fig:sim:tilting:lyapunov}, it can be seen that the positional Lyapunov candidate $V_\mathbf{x}$ is decreasing and converging, and the attitude converges to $\mathbf{R}_d$ within 3 seconds in stage B.
The control law succeeds to drive the vehicle to track the reference trajectory overall, despite slight tracking error in positions and velocities due to the discrete implementation of the continuous control law.
Meanwhile, the lack of angular velocity commands also leads to the delayed attitude response.

\subsection{Attitude Planner with Different Relaxation\label{sec:sim:tilting:relax}}
\begin{figure}[btp] 
    \centering
    \includegraphics[width=1.0\linewidth, trim={0 0 0 {0.5cm}}, clip]{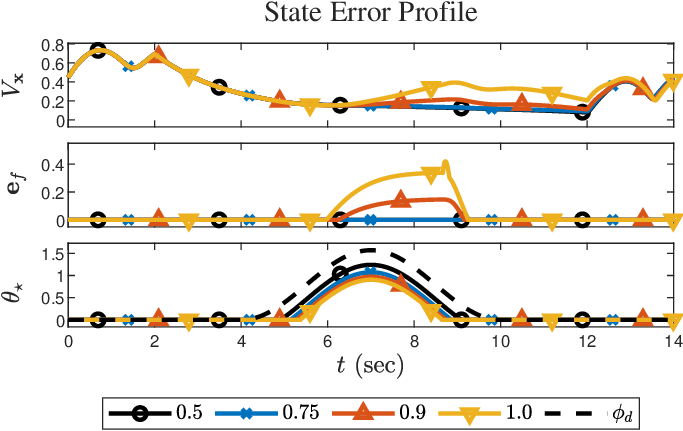}
\caption{The tracking performance with different attitude planner relaxtion, where the control force allocation error is defined as $\mathbf{e}_f = \|\mathbf{u}_{fd} - \mathbf{u}_{f}\|_2$ and $\mathbf{u}_{f}$ is the allocated control using redistributed pseudo inverse method \cite{jinModifiedPseudoinverseRedistribution2005}. $\phi_d$ is the reference attitude expressed in roll angle. \label{fig:sim:tilting:diff_s}}
\end{figure}

Since the control allocation problem is nonlinearly constrained, solutions may still be hard to obtain even though $\mathbf{u}_d \in \tilde{\mathcal{U}}_f$, especially when $\mathbf{u}_d$ is close to the AFS boundary.
Four different relaxations $s$ in attitude planner \eqref{eq:model:team:attainable_space_approx_semimajor_axis_x} and \eqref{eq:model:team:attainable_space_approx_semimajor_axis_y} are tested and the results are shown in \figref{fig:sim:tilting:diff_s}.
In the case of without any relaxation ($s=1.0$), the control force is not fully fulfilled when the vehicle is close to its maximum tilting angle at $t=6.2$, which leads to slightly drifting in positions. 
It also implies that the vehicle may entering infeasible attitude regions easily as encountering disturbances and model uncertainty.
If $s=0.5$ is used, the requested control $\mathbf{u}_d$ is far away to the AFS boundary, and the force allocation error $\mathbf{e}_f$ remains zero during the whole flight.
However, as $s$ decreases, the maximum tilting angle is sacrificed, which is a trade-off between tracking performance and robustness.

\section{Experiments}
A preliminary experiment is held with tethered power supply.
A Pixhawk (FMU-v2) serves as the main flight controller, running a customized PX4 autopilot to execute the full-pose control law and the control allocation algorithm for a TVMD vehicle frame. 
The Pixhawk sends the control commands of motors and servos of agents to the ESP32-based WiFi networked agent avionics.
The state feedback is provided by the EKF2 module in the PX4 autopilot with an additional onboard Jetson Nano running the Apriltag-ros package \cite{malyutaLongdurationFullyAutonomous2020} for the visual pose estimation of the team UAV.

\begin{figure}[btp] 
    \centering
    \includegraphics[width=1.0\linewidth, trim={0 0 0 0}, clip]{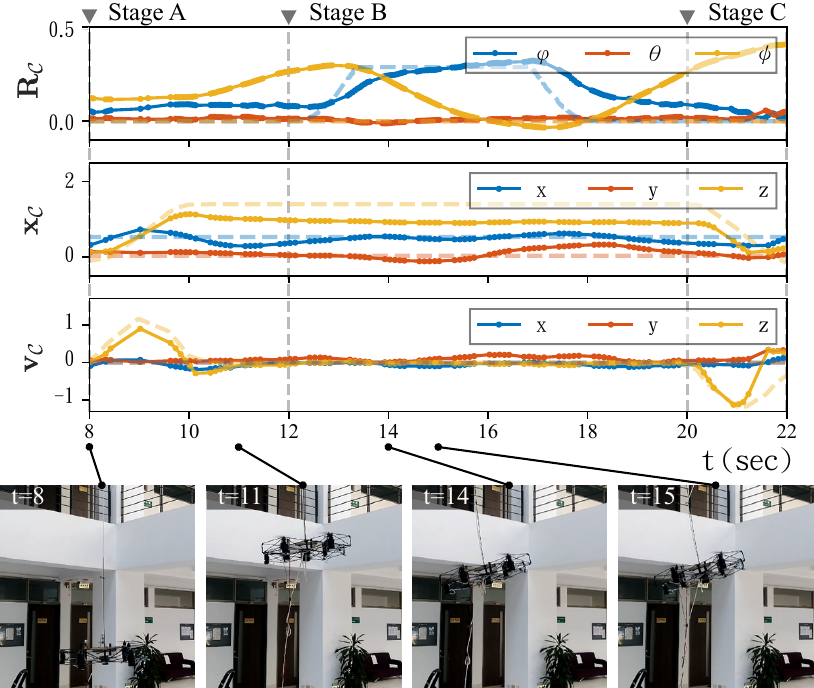}
\caption{The state profile in the experiment, where the dashed lines represent for the desird states, including $\mathbf{x}_r$, $\dot{\mathbf{x}}_r$, and the planned attitude $\mathbf{R}_d$. The attitude $\mathbf{R}_\mathcal{C}$ is expressed in Euler angles for easier understanding. The stick at the drone center is used to ensure the rope will not be tangled during the experiment, and the rope is loose after the vehicle ascends. \label{fig:exp:tilting:state}}
\end{figure}

\begin{figure}[btp] 
    \centering
    \includegraphics[width=0.9\linewidth, trim={0 0 0 0}, clip]{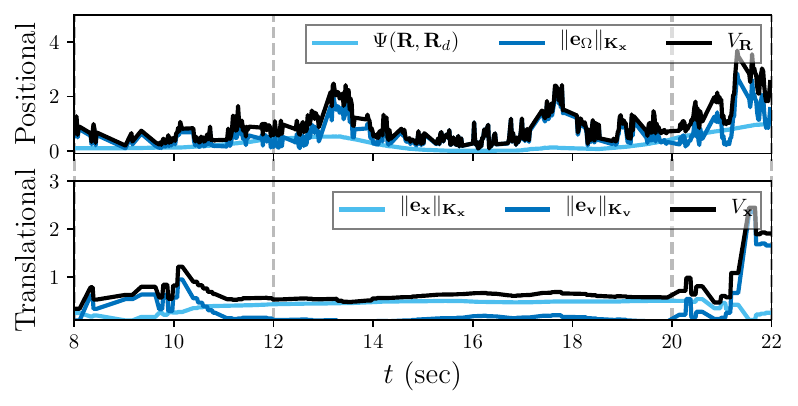}
\caption{The Lyapunov candidates profile in the experiment, where $V_\mathbf{R} = \Psi(\mathbf{R}_\mathcal{C}, \mathbf{R}_d) + \Vert\mathbf{e}_\Omega\Vert_{\mathbf{K}_\Omega}$ and $V_\mathbf{x} = \Vert\mathbf{e}_\mathbf{x}\Vert_{\mathbf{K}_\mathbf{x}} + \Vert\mathbf{e}_\mathbf{v}\Vert_{\mathbf{K}_\mathbf{v}}$. \label{fig:exp:tilting:lyapunov}}
\end{figure}

The maximum unallocated thrust control during the whole flight is less than $10^{-6}$ N.
The same reference trajectory, gains, and relaxations $s$ in \secref{sec:simulations} are used. \figref{fig:exp:tilting:state} shows the states and pictures at four time stamps.
Similar to the results in \figref{fig:sim:tilting:state}, $\mathbf{R}_d$ is truncated by the attitude planner \eqref{eq:ctrl:team:fullpose:attitude_planner} at about $\phi_d=0.28$ radian, which can be verified by re-calculating \eqref{eq:swarm:attainable_space_apprx}, and the roll angle $\phi$ tracks the desired attitude while holding its positions.
Steady z-axis position error appears, which is attributed to the weight of power cables and the model uncertainty of the thrust model \eqref{eq:model:thrust_force}.
Since the z-axis gain of $\mathbf{K}_\mathbf{R}$ is much smaller than the x- and y-axis, the attitude response in yaw angle drifted.
Other states have similar responses with the one in the simulation, which can also be seen from the Lyapunov candidates in \figref{fig:exp:tilting:lyapunov}.
After the vehicle ascends to the target height at $t=10$, the translational Lyapunov candidate $V_\mathbf{x}$ quickly drops to about 0.6 and remains at that level (which is mainly due to the z-axis steady state error) until entering stage C, and the attitude Lyapunov candidate $V_\mathbf{R}$ is also stabilized except for the accelerations during tilting at $t=13$ and $t=18$. 
Oscillations in positions can be also observed which is possibly due to the internal dynamics of servo motors; the delayed response may induce overshoots to the team system.


\section{Conclusions and Future Works}
A reconfigurable thrust-vectoring modular UAV is proposed which can achieve 6-DoF trajectory tracking.
%
With the attainable force space approximation and the cascaded control architecture, the control strategy can stabilize various team configurations without redesigning the controller.
%
The simulations show that the attitude planner generates feasible attitude targets according to the team configuration, and then the full-pose tracking controller stabilizes a 4-agent team system even encountering infeasible reference attitudes.
The experiment shows consistent results with the simulation. 
In future works, the model uncertainties and disturbances will be introduced, and the reference angular velocity will also be investigated.






\printbibliography

\end{document}